\documentclass[fleqn,usenatbib]{mnras}

\usepackage{graphicx}	
\usepackage{amsmath}	

\title[ Principal Component Analysis of Galaxy Clustering ]{ Principal Component Analysis of Galaxy Clustering in Hyperspace of Galaxy Properties }

\author[Zhou $\&$ Zhang]{
	Shuren Zhou,$^{1,2}$\thanks{E-mail: zhoushuren@sjtu.edu.cn}
	Pengjie Zhang, $^{1,3,2}$\thanks{E-mail: zhangpj@sjtu.edu.cn}
	Ziyang Chen $^{1,2}$
\\
$^{1}$Department of Astronomy, School of Physics and Astronomy,  Shanghai Jiao Tong University, Shanghai, China\\
$^{2}$Key Laboratory for Particle Astrophysics and Cosmology (MOE)/Shanghai Key Laboratory for Particle Physics and Cosmology, China\\
$^{3}$Division of Astronomy and Astrophysics, Tsung-Dao Lee Institute, Shanghai Jiao Tong University, Shanghai, China
}

\date{Accepted XXX. Received YYY; in original form ZZZ}

\pubyear{2022}

\begin{document}
\label{firstpage}
\pagerange{\pageref{firstpage}--\pageref{lastpage}}
\maketitle

\begin{abstract}
Ongoing and upcoming galaxy surveys are providing precision measurements of galaxy clustering. However a major obstacle in its cosmological application is the stochasticity in the galaxy bias. We explore whether the principal component analysis (PCA) of galaxy correlation matrix in hyperspace of galaxy properties (e.g. magnitude and color) can reveal further information on mitigating this issue. Based on the hydrodynamic simulation TNG300-1, we analyze the cross power spectrum matrix of galaxies in  the magnitude and color space of multiple photometric bands. (1) We find that the first principal component $E_i^{(1)}$ is an excellent proxy of the galaxy deterministic bias $b_{D}$, in that  $E_i^{(1)}=\sqrt{P_{mm}/\lambda^{(1)}}b_{D,i}$.  Here $i$ denotes the $i$-th galaxy sub-sample. $\lambda^{(1)}$ is the largest eigenvalue and $P_{mm}$ is the matter power spectrum. We verify that this relation holds for all the galaxy samples investigated, down to $k\sim 2h/$Mpc. Since $E_i^{(1)}$ is a direct observable, we can utilize it to design a linear weighting scheme to suppress the stochasticity in the galaxy-matter relation. For an LSST-like magnitude limit galaxy sample, the stochasticity $\mathcal{S}\equiv 1-r^2$ can be suppressed by a factor of $\ga 2$ at $k=1h/$Mpc. This reduces the stochasticity-induced systematic error in the matter power spectrum reconstruction combining galaxy clustering and galaxy-galaxy lensing from $\sim 12\%$ to $\sim 5\%$ at $k=1h/$Mpc. (2) We also find that $\mathcal{S}$ increases monotonically with $f_\lambda$ and $f_{\lambda^2}$. $f_{\lambda,\lambda^2}$  quantify the fractional contribution of other eigenmodes to the galaxy clustering and are direct observables. Therefore the two provide extra information on mitigating galaxy stochasticity.

\end{abstract}




\section{Introduction}
Galaxy clustering contains a wealth of cosmological information, in both its auto-correlation and cross-correlation with other fields of the large scale structure (LSS) of the Universe.  However, its application in precision cosmology is severely hindered by the complicated galaxy bias (\citet{desjacques2018large} for a recent review). The galaxy overdensity $\delta_g({\bf x})$ can be decomposed into a deterministic part and a stochastic part. In Fourier space, it reads
\begin{equation}
    \delta_g({\bf k})=b_D(k)\delta_m({\bf k})+\delta_g^S({\bf k})\ .
\end{equation}
Here $b_D(k)=P_{gm}(k)/P_{m}$ is the defined deterministic galaxy bias, and $\delta_m$ is the matter overdensity. The stochastic component $\delta_g^S$  includes both contribution from nonlinear and/or non-local $\delta_g$-$\delta_m$ relation, halo exclusion and shot noise caused by the discrete galaxy distribution \citep{1998ApJ...504..601P,1999ApJ...518L..69T,2004MNRAS.355..129S,bonoli2009halo,hamaus2010minimizing,2012PhRvD..85h3509C,baldauf2013halo,desjacques2018large,2019PhRvD..99l3514E}. With the presence of stochasticity, $b_D\neq b_S$ where $b_S\equiv \sqrt{P_{gg}/P_{mm}}$ is the other definition of galaxy bias.  Both definitions, $b_D$ and $b_S$, are widely used in the literature. We use the subscripts "$D$" and "$S$" to distinguish the two. The two are related to each other by $b_D=b_Sr$, where $r\equiv P_{gm}/\sqrt{P_{gg}P_{mm}}$ is the cross-correlation coefficient between the galaxy distribution and the matter distribution.  Precision modeling of $b_D$, $b_S$ and $r$ is highly challenging. 

Alternatively, if the deterministic bias $b_D$   is known or can be inferred to certain extent, new opportunities of cosmological applications can be realized.  One example is to suppress the stochasticity in $\delta_g$ by weighing halos with $b_D$ \citep{bonoli2009halo} or a function of host halo mass \citep{seljak2009suppress, hamaus2010minimizing, cai2011optimal, liu2021biased}. With reduced stochasticity, reconstruction of weak lensing maps \citep{ 2004MNRAS.350.1445P} and determination of structure growth rate with redshift space distortion \citep{2009JCAP...10..007M} can be achieved  beyond the cosmic variance limit. This will also enable accurate measurements of weak lensing power spectrum/correlation function combining galaxy-galaxy lensing and galaxy clustering \citep{2010PhRvD..81f3531B}, as well as reconstruction of dark matter power spectrum through analytic or semi-analytic models \citep{seljak2000analytic, guzik2001galaxy}. Another example is to weigh galaxies in different magnitude bins appropriately to reconstruct weak lensing by cosmic magnification \citep{2011MNRAS.415.3485Y,2015MNRAS.447..345Y}. A key condition here is to make the weighted $b_D$ vanishing. This condition requires to know the dependence of $b_D$ on magnitude. 

A major issue in reality is that observationally we are not able to directly measure $b_D$ nor individual host halo mass. Through principal component analysis (PCA), \citet{1999ApJ...518L..69T} postulated that the principal component, namely the largest eigenmode in the galaxy clustering matrix, is a good proxy of $r$.  \citet{bonoli2009halo} applied PCA to the halo clustering matrix in the halo mass space, confirmed the existence of the principal component and discovered the second eigenmode. Alternatively, \citet{hamaus2010minimizing} subtracted the deterministic halo clustering and focused on the eigenmode analysis of the remaining shot noise matrix in the halo mass space and found extra components.  

Here we focus on PCA of galaxy clustering matrix in space of observable properties such as magnitude and color, using the TNG300-1 simulation of the IllustrisTNG project \citep{Springel_2017, Nelson_2017, Pillepich_2017, Naiman_2018, Marinacci_2018}. The same simulation allows to directly measure $b_D$ and $r$ combining the simulated matter and galaxy distribution. So we can directly test the postulation 

The paper is organized as follows. In \S \ref{sec:methodology} we introduce the methodology of PCA and the postulated relation to connect direct observables with the desired properties ($b_D$ and $r$). In \S \ref{sec:data} we introduce the simulation and the galaxy samples to analyze. In \S \ref{sec:result} we present the major results. 

\begin{table*}
    \begin{tabular}{clc}
    \hline
      \bf{Galaxy samples}   &  \bf{Selection criteria} & $\bf{ \bar{n}_g \;[(Mpc/h)^{-3}] }$  \\
    \hline
      $S0$   & Full galaxy samples & $0.221 \;(z=0)$, $0.305 \;(z=1)$  \\
      $S1$   & Mass limitation ranging from $10^{9.5} M_{\odot}$ to $10^{11} M_{\odot}$ & $0.169 \;(z=0)$, $0.239 \;(z=1)$ \\
      $S2$   & Photometry $i$-band upper limitation as $-15$ Mag & $0.077 \;(z=0)$, $0.128 \;(z=1)$ \\
    \hline
      $S3$   & Flux limit with $griz$-bands cut at ($28.4, \,28.5, \,27.8, \,27.1$) Mag & $0.061 \;(z=1)$  \\
      $S4$   & Flux limit with $griz$-bands cut at ($27.4, \,27.5, \,26.8, \,26.1$) Mag & $0.043 \;(z=1)$  \\
      $S5$   & Flux limit with $griz$-bands cut at ($26.4, \,26.5, \,25.8, \,25.1$) Mag & $0.030 \;(z=1)$  \\
      $S6$   & Flux limit with $griz$-bands cut at ($25.4, \,25.5, \,24.8, \,24.1$) Mag & $0.020 \;(z=1)$  \\
    \hline
    \end{tabular}
    \caption{The galaxy samples used for the analysis. }
    \label{tab:sample}
\end{table*}

\section{Methodology}
\label{sec:methodology}
Galaxy clustering varies with observable galaxy properties such as magnitudes in multiple photometric bands and colors. So for a given survey and a given redshift range, we can split galaxies into sub-samples according to these properties. With $i=1,2\cdots, N$ sub-samples, we can measure the galaxy clustering matrix ${\bf P}$ at each $k$ bin. The matrix component $P_{ij}$ is the cross-power spectrum between the $i$-th and $j-$th galaxy sub-samples. If there is no stochasticity between matter and galaxies, $P_{ij}=b_{D,i}b_{D,j}P_{mm}$. Under this limit, the matrix ${\bf P}$ has only one eigenmode. Stochasticity in the galaxy-matter relation then naturally leads to extra eigenmodes in ${\bf P}$. Nevertheless, one may expect that the deterministic galaxy clustering is still dominant and remains as the principal component of ${\bf P}$. If so, we naturally expect 
\begin{equation}
\label{eqn:bdi}
    b_{D,i}\simeq \hat{b}_{D,i}\equiv\sqrt{\frac{\lambda^{(1)}}{P_{mm}}}E^{(1)}_i\ .
\end{equation}
Here the deterministic bias $b_D$ and $r$ of the $i$-the galaxy sub-samples are defined by
\begin{equation}
\label{eqn:bDid}
    b_{D,i}\equiv \frac{P_{im}}{P_{mm}}\ .
\end{equation}
$P_{mm}$ is the matter power spectrum, and $P_{im}$ is the cross power spectrum between the $i$-th galaxy sub-sample and the matter distribution. All the power spectra $P_{ij}$, $P_{mm}$ and $P_{im}$ are functions of $k$. This definition of bias (Eq. \ref{eqn:bDid}) differs from another widely used definition $\sqrt{P_{ii}/P_{mm}}$. The definition by Eq. \ref{eqn:bDid} only picks up the deterministic part of the bias, while the definition $\sqrt{P_{ii}/P_{mm}}$ contains the stochastic part of the bias. Therefore throughout the paper we adopt the bias definition of Eq. \ref{eqn:bDid}.
The $E^{(1)}_i$ is the $i$-th element of the first eigenvector of the matrix ${\bf P}$. Specifically, we decompose ${\bf P}$ into its eigenmodes of eigenvalue $\lambda^{(\alpha)}$ and eigenvector ${\bf E}^{(\alpha)}$. By the spectral decomposition theorem, 
\begin{equation}
    P_{ij}=\sum_\alpha \lambda^{(\alpha)} E^{(\alpha)}_i E^{(\alpha)}_j\ .
\end{equation}
We adopt the convention $\lambda^{(1)}\geq \lambda^{(2)}\geq \cdots$, so $\lambda^{(1)}$ is the eigenvalue of the principal component. 
Eq. \ref{eqn:bdi} is equivalent to what has been proposed by \citet{1999ApJ...518L..69T}, 
\begin{equation}
\label{eqn:ri}
    r_i\simeq \hat{r}_i\equiv\sqrt{\frac{\lambda^{(1)}}{P_{ii}}}E^{(1)}_i\ .
\end{equation} 
Here the cross-correlation coefficient $r$ is defined by
\begin{equation}
 r_i\equiv \frac{P_{im}}{\sqrt{P_{ii}P_{mm}}}\ .
\end{equation}
Correspondingly we can define the stochasticity as 
\begin{equation}
    \mathcal{S}\equiv 1-r^2\ .
\end{equation}
The r.h.s. of Eq. \ref{eqn:bdi} \& \ref{eqn:ri} are obtained from the simulation data. Eq. \ref{eqn:bdi} requires a fiducial cosmology to predict $P_{mm}$, while Eq. \ref{eqn:ri} has the advantage of the completely determined from observations.  

If Eq. \ref{eqn:bdi} \& \ref{eqn:ri} hold, we can directly infer $b_D$ and $r$ from observations and use them to enhance cosmological applications of galaxy clustering. A straightforward application is to reconstruct $P_{mm}$, combining $P_{im}$ available from galaxy-galaxy lensing and $P_{ij}$. For each pair of $i$-th and $j$-th sub-sample, we can build an estimator of $P_{mm}$
\begin{equation}
	\hat{P}_{mm} \,=\, \frac{ P_{im}P_{jm} }{ \lambda^{(1)} E^{(1)}_{i} E^{(1)}_{j} }\ .
\end{equation}
This turns out to be
\begin{equation}
    \hat{P}_{mm}=\left(\frac{b_{D,i}}{\hat{b}_{D,i}}\right)\left(\frac{b_{D,j}}{\hat{b}_{D,j}}\right) P_{mm}=\left(\frac{r_i}{\hat{r}_i}\right)\left(\frac{r_j}{\hat{r}_j}\right) P_{mm}\ .
\end{equation}
Namely the accuracy of Eq. \ref{eqn:bdi} \& \ref{eqn:ri} determines the accuracy of $\hat{P}_{mm}$. 

Since there are totally $N(N+1)/2$ independent pairs of $P_{im}$ and $P_{jm}$, we should combine all of them to further improve the $P_{mm}$ reconstruction. We seek for the following weighting scheme, 
\begin{equation}
\hat{P}_{mm} \,=\, \frac{\sum_{ij} W_{ij} P_{i m}P_{jm}}{\sum_{ij} W_{ij}\lambda^{(1)}  E^{(1)}_i  E^{(1)}_j}\ .
\end{equation}
The weight $W$ which extremizes $\hat{P}_{mm}$ is
\begin{equation}
W_{ij}\propto E^{(1)}_i  E^{(1)}_j\ .
\end{equation}
Therefore
\begin{equation}
\label{eqn:Pmm_weighted}
   \hat{P}_{mm} \,=\, \frac{(\sum_i  E^{(1)}_i P_{i m})^2}{\lambda^{(1)} } \ .
\end{equation}
Here we have used the eigenvector normalization condition, $\sum_i^{N}{(E_i^{(1)})^2}=1$.  Another derivation of this weighted estimator is presented in the appendix.~\ref{append:another_derivation}.

There are also other measures of stochasticity, motivated by the fact that the stochasticity vanishes in the limit of  $\lambda^{(i\geq 2)}=0$.   Since we have the relation
\begin{equation}
\sum_\alpha \lambda^{(\alpha)}=\sum_i P_{ii}\ ,    
\end{equation}
the property $f_\lambda$ quantifies the contribution of $i=2,3,\cdots$ eigenmodes to $P_{ii}$ and therefore provides a measure on the overall stochasticity. Here
\begin{equation}
\label{eqn:flambda}
    f_{\lambda}\equiv 1-\frac{\lambda^{(1)}}{\sum_\alpha \lambda^{(\alpha)}}\ .
\end{equation}
We also have the relation
\begin{equation}
    \sum_\alpha (\lambda^{(\alpha)})^2=\sum_{ij} P_{ij}^2\ .
\end{equation}
Therefore we can define $f_{\lambda^2}$ as another measure of stochasticity. 
\begin{equation}
\label{eqn:flambda2}
    f_{\lambda^2}\equiv 1-\frac{(\lambda^{(1)})^2}{\sum_\alpha (\lambda^{(\alpha)})^2}\ .
\end{equation}

The bottom line is that through PCA of the directly observed galaxy clustering matrix, we could gain information on the otherwise unknown stochasticity between the galaxy distribution and matter distribution. Now we use simulations to test the accuracy of Eq. \ref{eqn:bdi}, to explore the $f_{\lambda, \lambda^2}$-$\mathcal{S}$ relation and to quantify such gain in information.

\begin{figure*}
	\includegraphics[width=0.80\textwidth]{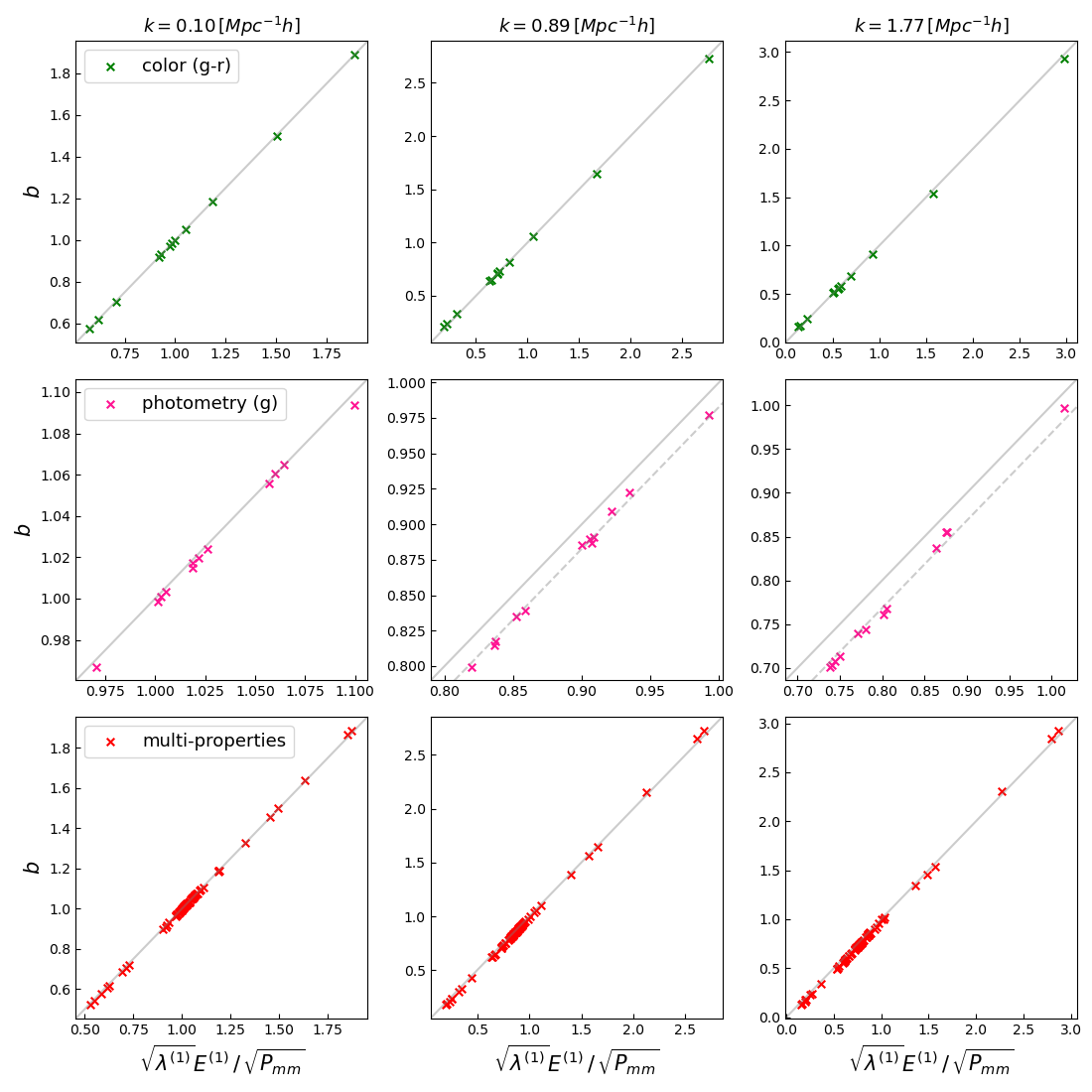}
    \caption{Numerical verification of Eq. \ref{eqn:bdi}, namely $b_{D,i}=\sqrt{\lambda^{(1)}}E_i^{(1)}/\sqrt{P_{mm}}$ (the diagonal lines), for the galaxy sample $S0$. Here $i\in (1,N)$ denotes the $i$-th sub-sample. $b_{D,i}$ is the corresponding deterministic bias measured directly through the simulated matter and galaxy distribution. $\lambda^{(1)}$ is the largest eigenvalue of the $N\times N$ cross power spectrum matrix, and $E^{(1)}$ is the corresponding eigenvector. For brevity, we only show the case in the $g-r$ color space ($N=12$), $g$-band magnitude space ($N=12$) and the hyperspace of 3 color and 4 photometric bands ($N=84$). We show the results at $k=0.10\,{\rm Mpc}^{-1}h\, ,0.89\, {\rm Mpc}^{-1}h \,$ and $1.77\, {\rm Mpc}^{-1}h$ from the left panels to the right panels respectively.  Notice the difference in the range of vertical axis, which highlights the derivation to Eq. \ref{eqn:bdi} in the galaxy photometry space compared to that in the color space and multi-property hyperspace. The offset to Eq. \ref{eqn:bdi}, shown as the dashed lines in the middle panels, is $2\%$ at  $k=0.89[Mpc^{-1}h]$ and $4\%$ at $k=1.77[Mpc^{-1}h]$.}
    \label{fig:5}
\end{figure*}
\section{Data}
\label{sec:data}

We utilize the publicly available TNG300-1 simulation from the IllustrisTNG project \citep{Springel_2017, Nelson_2017, Pillepich_2017, Naiman_2018, Marinacci_2018}.
IllustrisTNG is a suite of large volume, cosmological, gravo-magnetohydrodynamical simulations run with the moving-mesh code ${\tt APEPO}$ \citep{springel2010pur}.
TNG implements a set of physical processes to model the galaxy formation process including gas cooling, star formation and evolution, supernova feedback \citep{pillepich2018simulating},
 and AGN feedback \citep{weinberger2016simulating}
This project includes TNG50, TNG100 and TNG300, with boxsize of $50$ Mpc, $100$ Mpc and $300$ Mpc respectively. We choose TNG300 for our analysis, for its largest simulation volume and galaxy sample. 


The TNG galaxy catalog provides many galaxy properties. But most of them, including galaxy stellar and halo mass, are not direct observables. We choose galaxy magnitudes in four photometry bands ($griz$, \citet{2002AJ....123..485S}) and three galaxy colors ($g$-$r$, $r$-$i$, $i$-$z$) as the primary observed galaxy properties. For simplicity, we do not consider various observational issues such as dust extinction and correction, and redshift errors.  

The full galaxy sample of TNG300-1 simulation contains  $1.9\times10^6$ galaxy objects at $z=0$ and $2.6\times10^6$ galaxy objects at $z=1$, with the mean galaxy number density $\bar{n}_g=0.221 (\rm Mpc/h)^{-3}$ and $\bar{n}_g=0.305 (\rm Mpc/h)^{-3}$ at $z=0$ and $z=1$ respectively. 

We will perform the PCA analysis on this galaxy sample ($S0$), at both $z=0$ and $z=1$. In order to examine the generality and feasibility of Eq. \ref{eqn:bdi}, we apply various cuts to form the other 6 samples, listed in Table \ref{tab:sample}.  For the flux-limited galaxy samples in the Table \ref{tab:sample}, S4 takes the magnitude cuts corresponding to the expected depth of coadded images from LSST \citep{ivezic2019lsst}. We also increase or decrease the magnitude cuts to generate the other 3 flux-limited samples, $S3$, $S5$ and $S6$.

For the cross power spectrum matrix measurement of each galaxy sample, we split the selected samples into $N=12$ sub-samples according to each given galaxy property, i.e. photometry or color. The number of galaxies in each sub-sample bin is equal, for homogeneous control over the impact of shot noise.  We label the combination result using all the sub-samples with the name multi-properties below, in which totally 84 sub-samples comes from 7 kinds of properties times 12 sub-samples in each property hyperspace.
To examine the robustness of our binning strategy, we also compute the weighted estimator with $N=20$ sub-samples bins, under which we have the correlation matrix with dimension $20\times20$ in each galaxy hyperspace and $140\times140$ in the combined galaxy property hyperspace. The final weighted estimator with $N=20$ presents a slight difference compared to those with $N=12$, so our choice of $N=12$ is numerically stable. Therefore for brevity, we do not show the results of $N=20$ in this paper.

\section{Results}
\label{sec:result}
We utilize the package ${\tt Nbodykit}$ \citep{Hand_2018} to compute the power spectrum, and set the mesh grid number as $600^3$. Given the relatively small boxsize, we will focus on the $k$ range of $0.1h/$Mpc$\leq k\leq 2.4h/$Mpc.

\subsection{Testing Eq. \ref{eqn:bdi}}
\begin{figure*}
	\includegraphics[width=0.75\textwidth]{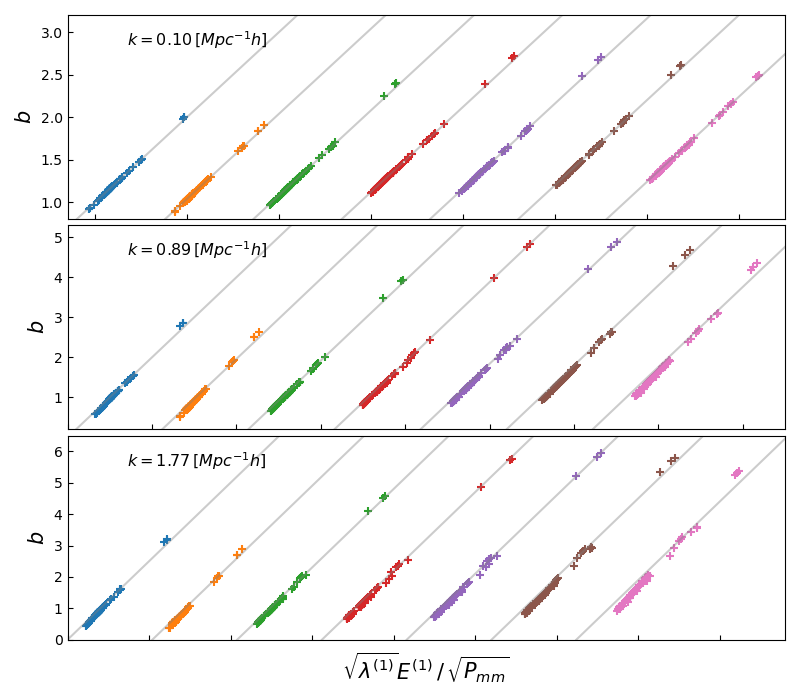}
    \caption{Numerical verification of Eq. \ref{eqn:bdi}, for all 7 galaxy samples ($S0$, $S1$, $S2$, $S3$, $S4$, $S5$ and $S6$). For clarity, we  shift the data points of $S0$-$S6$ horizontally from left to right, while the sold straight lines are the prediction of Eq. \ref{eqn:bdi}. This relation holds regardless of galaxy samples and galaxy properties.  }
    \label{fig:6}
\end{figure*}

Fig. \ref{fig:5} shows the test against Eq. \ref{eqn:bdi} for the $S0$ galaxy sample. The relation holds for scales investigated ($k=0.1h/$Mpc to $k=1.77h/$Mpc). It also holds in the color space, magnitude space of a single photometric band, or higher dimensional spaces of color and photometry combination. Eq. \ref{eqn:bdi} also holds for other galaxy samples, as shown in Fig. \ref{fig:6}. 
Because of the narrow range of $b_D$ spanned in the photometric space, there is a visually significant offset from equality. We suppose there is residual stochasticity in the principal component in photometric space, and this part of stochasticity causes the ensemble over-estimation of the $b_D$. The nearly constant offset implies the residual stochasticity is dominated by equally contaminating shotnoise, since we divide galaxies into bins of equal number.

\begin{figure}
	\includegraphics[width=\columnwidth]{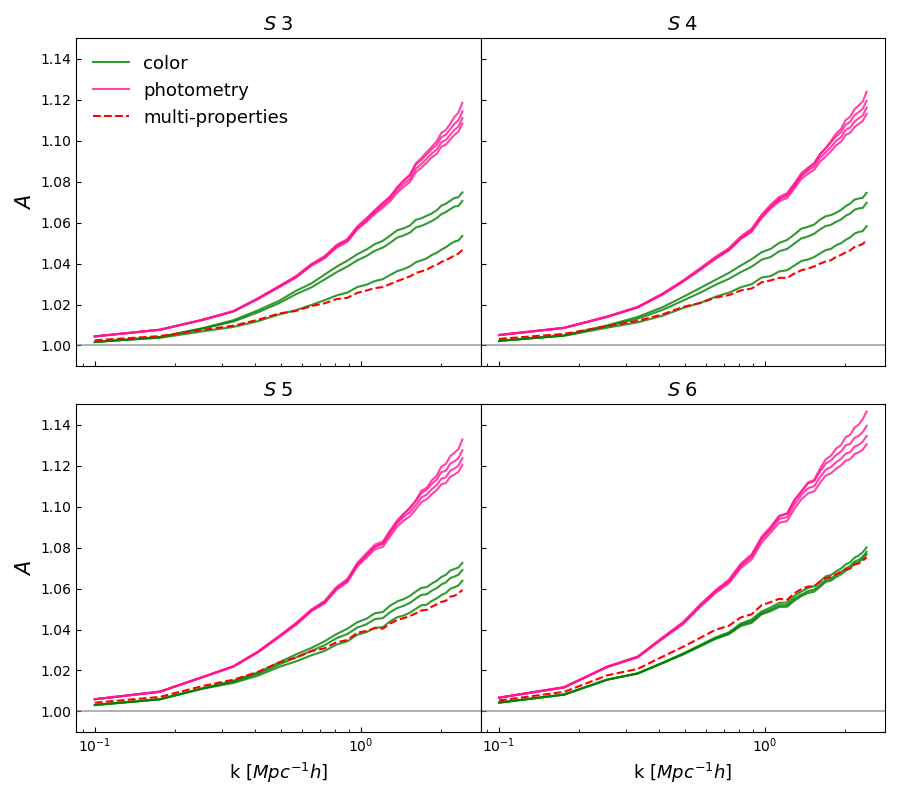}
    \caption{ The overall amplitude $A$, defined as the bestfit parameter of relation $\hat{b}_{D,i} = A\,b_{D,i}$, as a function of scale $k$. The $A$ deviates from unity at the level of averagely $5\%$ at $k=1h/$Mpc, for the galaxy sample $S3$-$S6$. The case in hyperspace of 3 color and 4 photometry combination presents small deviation compared to those in single color /photometry hyperspace, and similar phenomena appear in $S0$-$S2$. } 
    \label{fig:77}
\end{figure}

We quantify the accuracy of Eq. \ref{eqn:bdi} with two measures. One is the overall amplitude $A$ and the other is the cross-correlation coefficient $R$ (Pearson's correlation coefficient). $A$ is  defined by fitting $\hat{b}_{D,i}$ with $A\,b_{D,i}$. The bestfit of $A$ is
\begin{equation}
    A=\frac{\sum_i b_{D,i}\hat{b}_{D,i}}{\sum_i b_{D,i}^2}\ .
\end{equation}
Fig. \ref{fig:77} shows $A$ as a function of $k$, for the galaxy samples $S3$-$S6$. At $k<1h/$Mpc, the deviation of $A$ from unity is less than $10\%$, in the photometry space. The deviation is reduced to $5\%$ or less in the color space. The deviation from unity is usually the smallest in the  color-photometry space. 

\begin{figure}
	\includegraphics[width=\columnwidth]{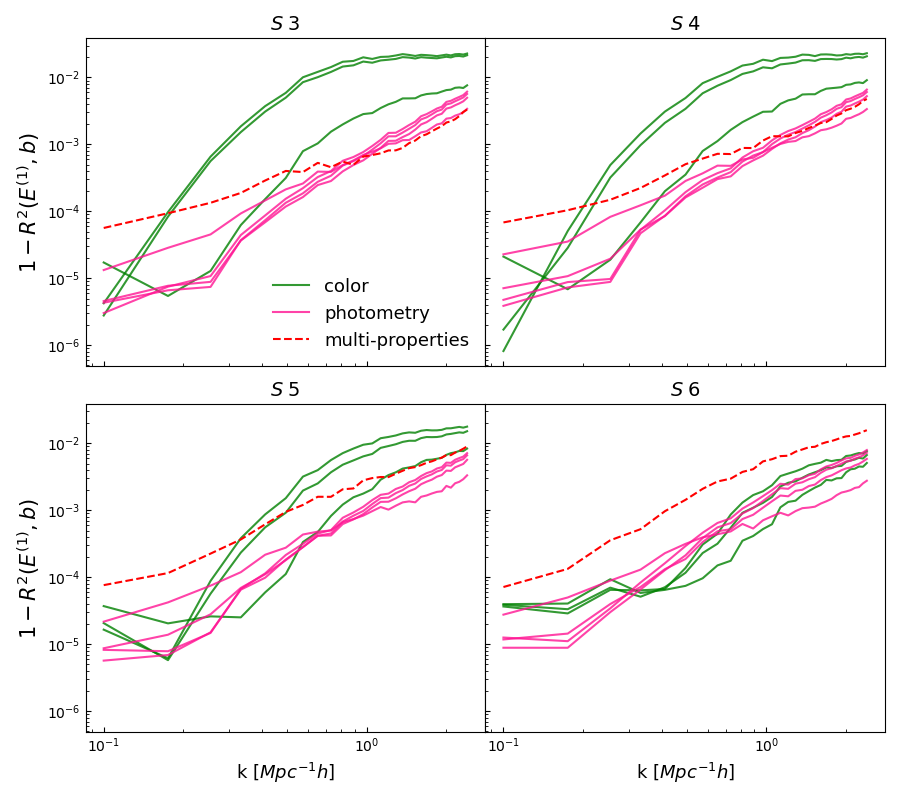}
    \caption{The proportionality between $b_{D,i}$ and $E^{(1)}_i$, quantified by the cross correlation coefficient $R$ as a function of scale $k$. $R$ deviates from unity at the level of $10^{-2}$ at $k=1h/$Mpc, for the galaxy sample $S3$-$S6$. The deviation from unity is even smaller for $S0$-$S2$. $R$ in the photometry space is usually closer to unity than that in the color space.  }
    \label{fig:7}
\end{figure}

In some applications, the accuracy and the overall amplitude $A$ for Eq. \ref{eqn:bdi} is irrelevant, and what matters is the proportionality relation. In such case, the accuracy of this relation is quantified by the measurement $R$, 
\begin{equation}
	R(E^{(1)}, b) = \frac{ \sum_i{b_{D,i}\hat{b}_{D,i}}}{ \sqrt{ \sum_i{b_{D,i}^2} \sum_i{ \hat{b}_{D,i}^2 } } }=\frac{ \sum_i{b_{D,i}E_i^{(1)}} }{ \sqrt{ \sum_i{b_{D,i}^2} \sum_i{ (E_i^{(1)})^2 } } }\ .
\label{equ:correlation_R}
\end{equation}
Fig.~\ref{fig:7} shows $1-R^2(E^{(1)}, b)$ in the color space, and in the magnitude space of each photometric bands. This value is $\la 10^{-2}$ at $k<1h/$Mpc and shows no significant increase at $k>1h/$Mpc. The deviation of $R$ from unity is $<1\%$ over all scales investigated. Therefore  $R(E^{(1)}, b)=1$ is essentially exact, namely $b_{D,i}\propto E^{(1)}_i$ is nearly exact. 
The measurement of $A$ shows that the performance in color space is better than in the photometric space, while the measurement of $R(E^{(1)}, b)$ seems to present a contrary conclusion. The reason is that the overall amplitude $A$ quantifies the systematic tendency of $E^{(1)}$ and $b_{D}$, even there is scattering. However, any scattering will reduce the correlation, therefore reduce the value of $R(E^{(1)}, b)$, since the fluctuation in $E^{(1)}$ enter the denominator of Eq. \ref{equ:correlation_R} in the form of square and are unable to cancel each others. 
In photometric space, the mildly ensemble offset of $E^{(1)} - b_D$ equality dominates the deviation of $A$ from $1$, results in the poorest performance in the photometric space. When it comes to the measurement of $R(E^{(1)}, b)$, the fluctuations in the color space overcome the slightly systematic offset in photometric space, so the performance in color space is worst.

We caution that the above findings are based on a single simulation (and refer to \citet{nelson2019illustristng} for further discussion of the TNG simulations). Nevertheless we expect that Eq. \ref{eqn:bdi} is in general valid and its accuracy in general improves with increasing dimension of galaxy property hyperspace. The reason is that the deterministic clustering is universally presented in all galaxy sub-samples, while the stochastic clustering is not. Therefore including more galaxy sub-samples in the cross power spectrum matrix $P_{ij}$ enhances the contribution of deterministic clustering, and improves the accuracy of Eq. \ref{eqn:bdi}. 

\subsection{Testing the matter power spectrum reconstruction accuracy}
After validating Eq. \ref{eqn:bdi}, we proceed to quantify the accuracy of the matter power spectrum $\hat{P}_{mm}$  reconstructed through Eq. \ref{eqn:Pmm_weighted}. The benchmark to compare is the most straightforward reconstruction $P_{gm}^2/P_{gg}=r^2P_{mm}$, which has a systematic error $\mathcal{S}=1-r^2$. So improving the reconstruction accuracy is equivalent to suppressing the stochasticity. 

\begin{figure*}
	\includegraphics[width=0.96\textwidth]{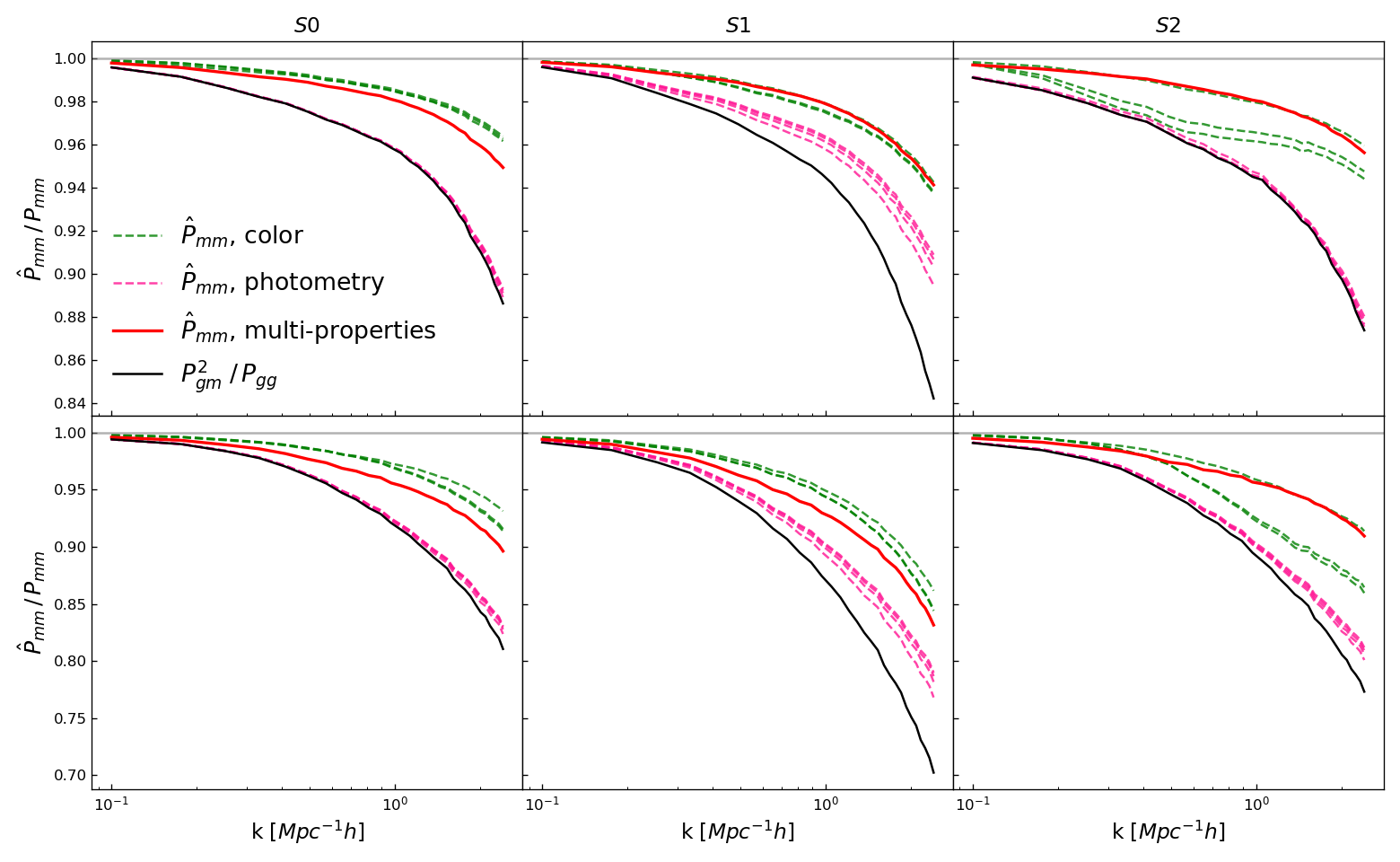}
    \caption{  The ratio of reconstructed matter power spectrum $\hat{P}_{mm}$ to the $P_{mm}$ measured directly from simulation, for samples $S0$-$S2$ from left to right. The top/bottom panels show the snapshot at redshift z=0/z=1 respectively. Black solid line present ratio of the benchmark reconstruction $P_{gm}^2/P_{gg}$ to $P_{mm}$. Green-dash, pink-dash and red-solid lines are reconstructed $\hat{P}_{mm}$ through Eq. \ref{eqn:Pmm_weighted} in galaxy color, photometry and color-photometry hyperspace. }
    \label{fig:8}
\end{figure*}

Fig.~\ref{fig:8} shows the ratio $\hat{P}_{mm}/P_{mm}$ for galaxy samples $S0$, $S1$ and $S2$. Here the true $P_{mm}$ is measured directly from the simulation,  The weighted estimator of Eq. \ref{eqn:Pmm_weighted} indeed suppresses stochasticity and improves the power spectrum reconstruction. For the $S0$ sample with $z=0$ in the galaxy color space, the improvement in the reconstruction accuracy is  $4.3\%\rightarrow 2.0\%$ at $k=1h/$Mpc. For $S1$($S2$), the improvement is $5.6\%\rightarrow 2.1\%$ ($5.6\%\rightarrow 2.0\%$) at $k=1h/$Mpc. The improvement at $z=1$ is similar.  We also notice that the improvement depends on the hyperspace to perform the PCA analysis. The improvement in the magnitude space is negligible for the $S0$($S2$) sample. For the $S1$ sample, the improvement is visible, but not as significant as in the color space. Nevertheless, the PCA analysis in the higher dimensional color-magnitude hyperspace is always efficient in improving the $P_{mm}$ reconstruction.

\begin{figure*}
	\includegraphics[width=0.75\textwidth]{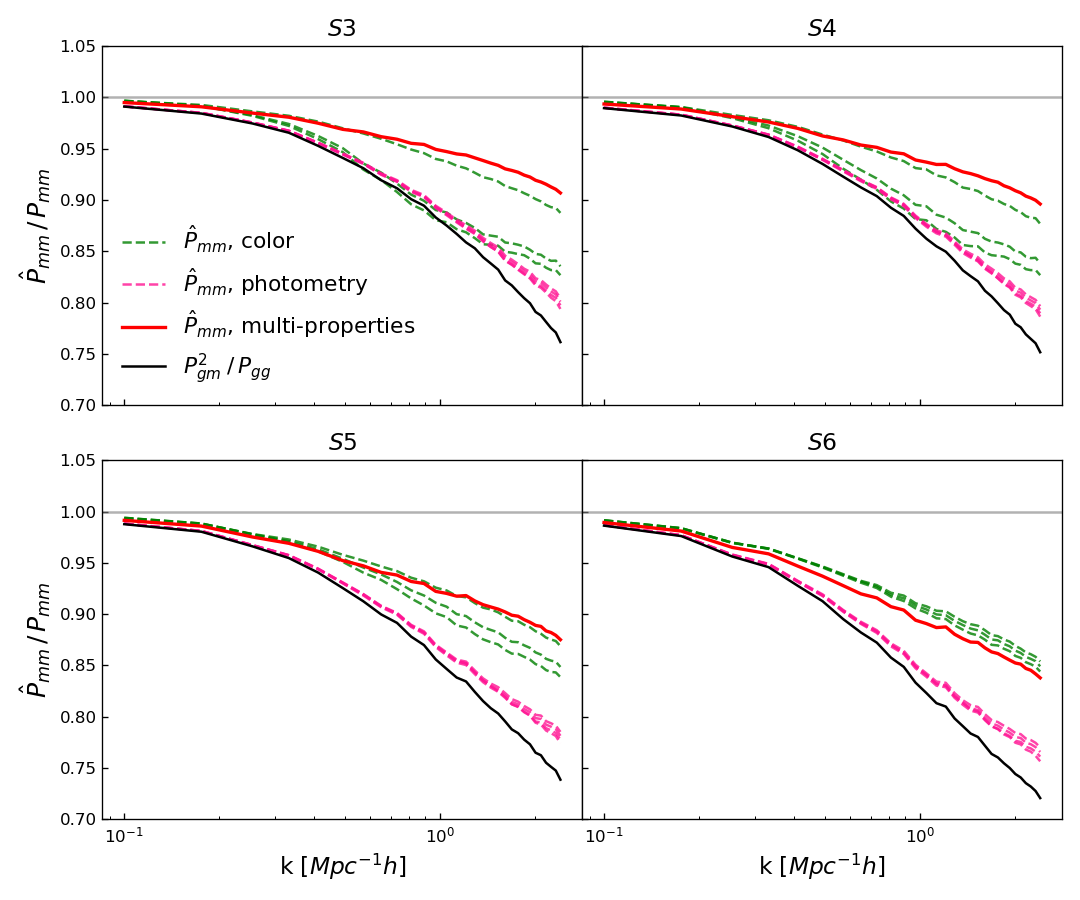}
    \caption{ Same figure as Fig.\ref{fig:8}, but present the galaxy sample $S3$-$S6$. }
    \label{fig:9}
\end{figure*}

Fig.~\ref{fig:9} shows the results for the galaxy samples $S3$-$S6$. The $\hat{P}_{mm}$ improvement shows more diversified dependence on the hyperspace to perform the reconstruction. The improvement is usually negligible in the photometry space, no matter which photometric band is used. The improvement in the color space is significant in one color space (color $i-z$ in our results), but not in another color space. Nevertheless, the improvement in the combined color-photometry hyperspace is always significant. For S4 (LSST-like magnitude limit), the reduction in systematic error of reconstruction is $11.7\%\rightarrow 5.1\%$ at $k=1h/$Mpc. This improvement is insensitive to the magnitude limit. For brighter galaxy samples S5(S6), the reduction is  $12.8\%\rightarrow 5.9\%$($14.4\%\rightarrow 7.0\%$). For fainter galaxy sample S3, the performance is similar. 

The difference in the performance of different galaxy hyperspace is determined by the dynamical range of $b_{D}$.  Across the photometry space, $b_{D}$ varies only by $\sim 10-30\%$ (Fig. \ref{fig:5} \& \ref{fig:6}). Therefore the weights in Eq. \ref{eqn:Pmm_weighted} for different sub-samples are essentially the same. For this reason, this weighting scheme does not show significant improvement of power spectrum reconstruction. In contrast, $b_{D}$ across the color space (and the color-photometry space) varies by a factor of $2$ or more, making the weighting scheme more efficient. Since observationally we know the dynamical range of $b_D$ through $\hat{b}_D$, we can always choose galaxy sub-samples of larger dynamical range in order to have better $P_{mm}$ reconstruction. 

\subsection{$f_\lambda$-$\mathcal{S}$ relation}
\begin{figure*}
	\includegraphics[width=0.75\textwidth]{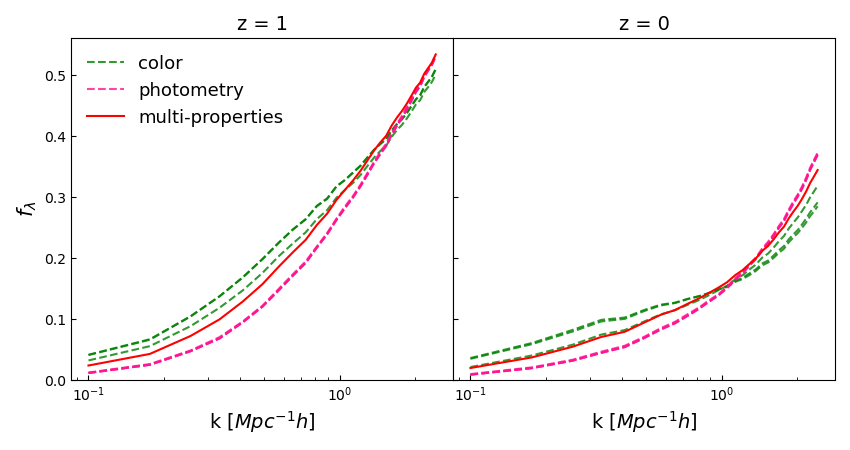}
    \caption{ $f_\lambda$ as a function of scale $k$ for samples $S0$. The left/right panel present the snapshot at redshift $z=1$/$z=0$ respectively. Green-dash, pink-dash and red-solid lines are results in galaxy color, photometry and color-photometry hyperspace. }
    \label{fig:12}
\end{figure*}

\begin{figure*}
	\includegraphics[width=0.80\textwidth]{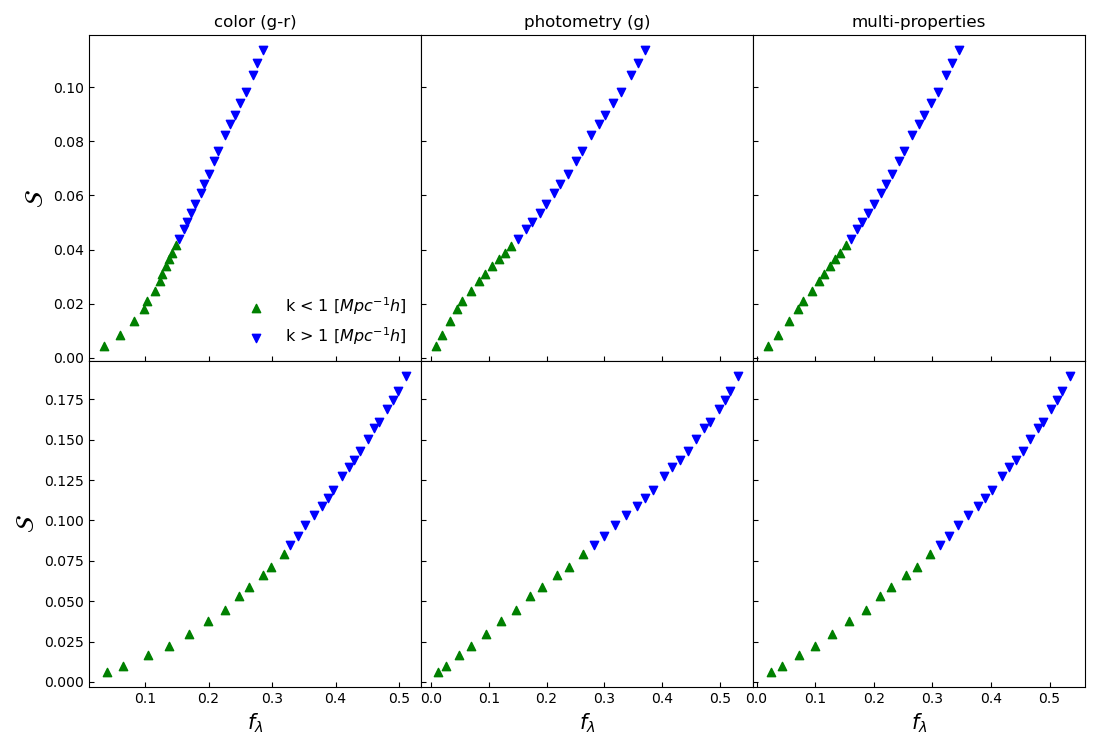}
	\includegraphics[width=0.80\textwidth]{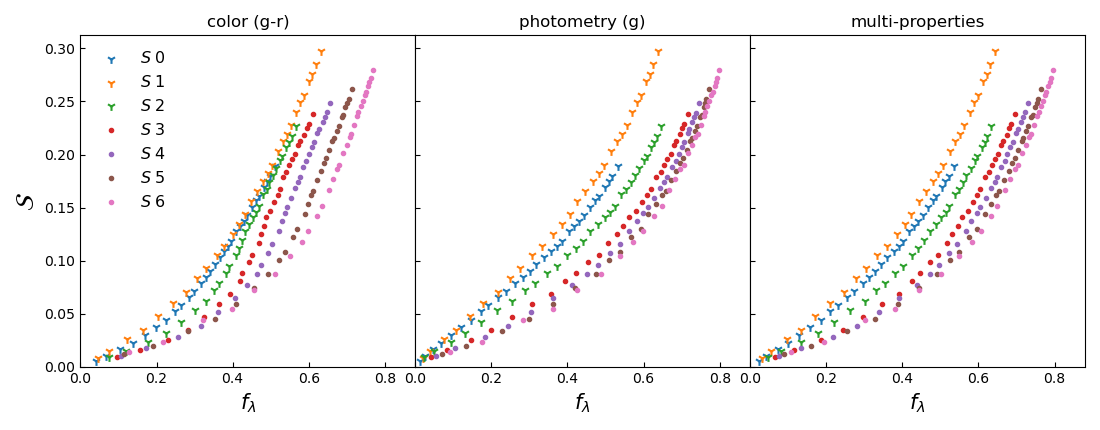}
	\includegraphics[width=0.80\textwidth]{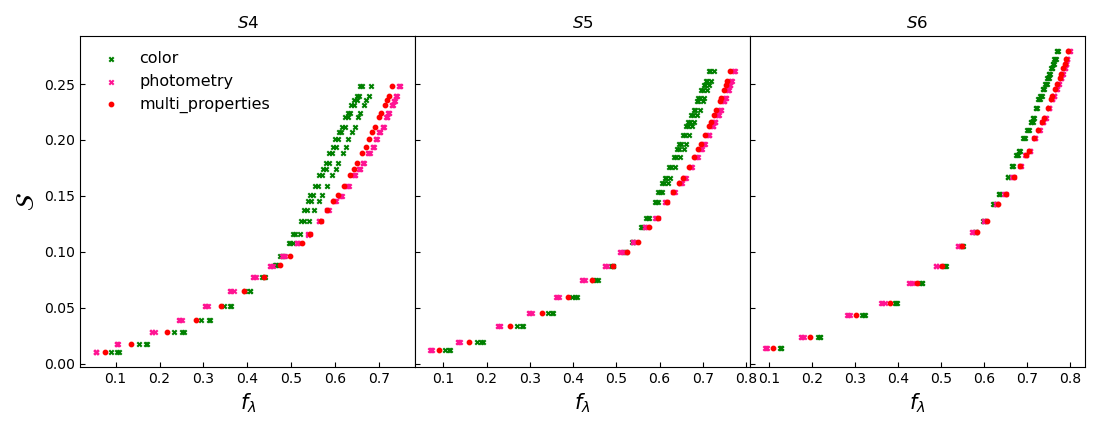}
    \caption{ The relation between $f_{\lambda}$ and $\mathcal{S}=1-r^2$. The first and second rows show the results for galaxy samples $S0$ at redshift $z=0$ and $z=1$ respectively. Different points represent different $k$ bins and the $k$ value larger than $1\,{Mpc}^{-1}h$ or smaller than $1\,{Mpc}^{-1}h$ is marked by different markers. The third row show the results for different galaxy sample selections at $z=1$. Column panels show sub-samples divided by galaxy color(g-r), galaxy photometry(g) or multi-properties. 
	The last row show the $f_{\lambda} - \mathcal{S}$ relation for the various galaxy binned sub-samples for various flux limited sample, $S4$, $S5$ and $S6$ from left to right. }
    \label{fig:34}
\end{figure*}

$f_{\lambda}$ (Eq. \ref{eqn:flambda}) and $f_{\lambda^2}$  (Eq. \ref{eqn:flambda2}) are also observables. So in principle useful information on stochasticity can be inferred from both measures.  Fig. \ref{fig:12} shows $f_\lambda$ as a function of $k$ for the S0 sample. Contributions from eigenmodes other than the largest one increase towards small scales. $f_\lambda$ reaches $20$-$30\%$ at $k=1h/$Mpc. Although this is not directly related to the stochasticity between galaxy and matter, it is indeed tightly related, as shown in Fig. \ref{fig:34}. The stochasticity $\mathcal{S}=1-r^2$ increases monotonically with $f_\lambda$, regardless of galaxy sample, galaxy property hyperspace and $k$ range. 

The $\mathcal{S}$-$f_\lambda$ relation shows significant scatter across the galaxy samples ($S0$-$S6$, middle panel, Fig. \ref{fig:34}). Within each galaxy sample, the scatter across different galaxy properties hyperspace is significantly smaller (bottom panel, Fig. \ref{fig:34}). 

\begin{figure*}
	\includegraphics[width=0.80\textwidth]{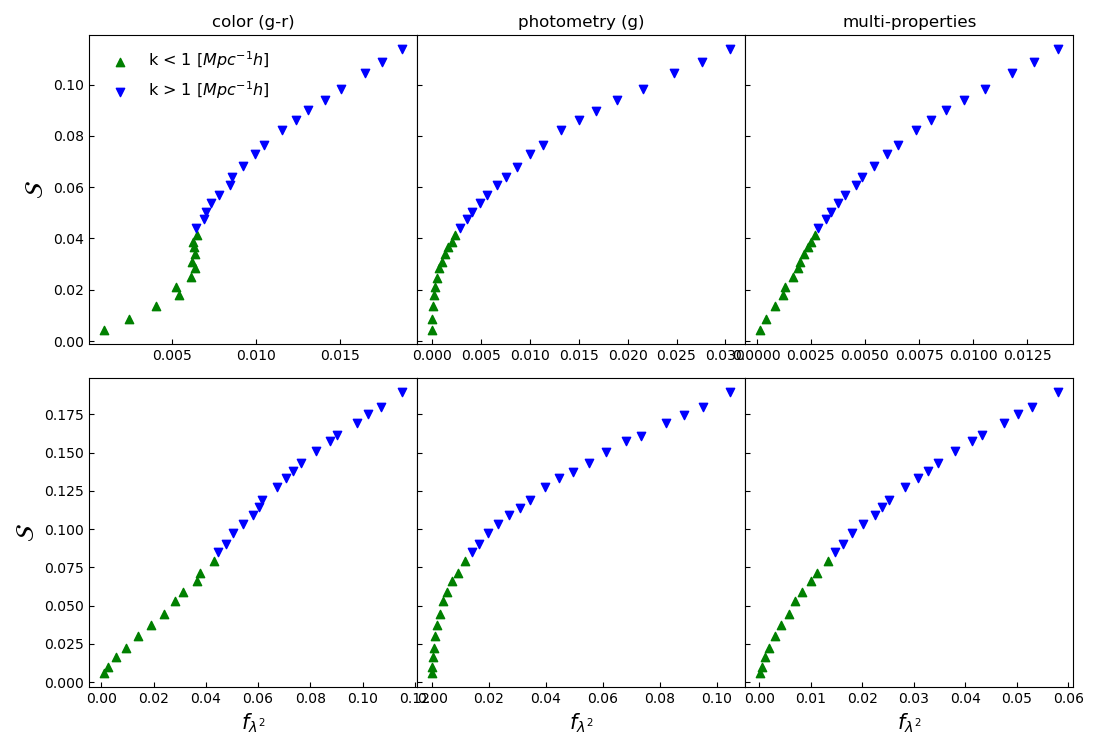}
	\includegraphics[width=0.80\textwidth]{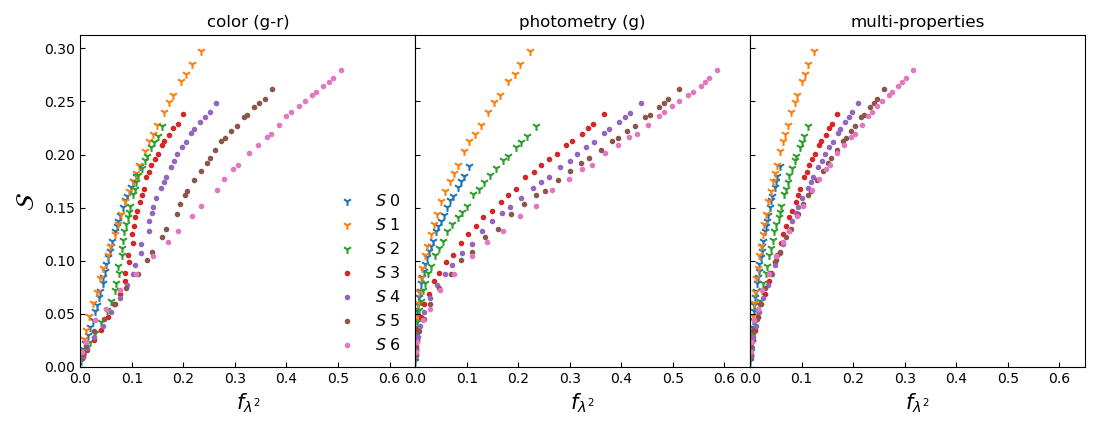}
	\includegraphics[width=0.80\textwidth]{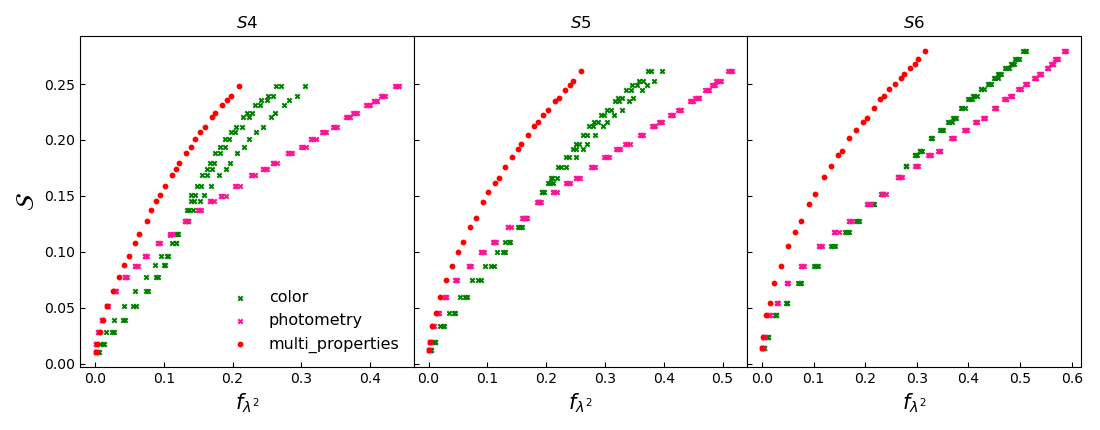}
    \caption{ Same figure as Fig.~\ref{fig:34}, but present the relation $f_{\lambda^2}$ - $\mathcal{S}$. }
    \label{fig:34_more}
\end{figure*}

Scatters in the $\mathcal{S}$-$f_\lambda$ relation mean extra factors affecting $\mathcal{S}$. As a first attempt, we check $f_{\lambda^2}$. The reason is that $f_{\lambda^2}$ is more sensitive to the second largest eigenmode than $f_{\lambda}$. Fig. \ref{fig:34_more} shows the $\mathcal{S}$-$f_{\lambda^2}$ plot. It indeed shows different scatter patterns,  in particular at $\mathcal{S}\ga 0.2$. This implies that a combination of $f_\lambda$ and $f_{\lambda^2}$ may form a tighter relation with $\mathcal{S}$. This possibility, along with other possible measures of stochasticity using the matrix $P_{ij}$, are worthy of further investigation. The bottom line is that through the galaxy matrix analysis, we can form various observables, which contains information on the galaxy-matter stochasticity. A thorough investigation is beyond the scope of this paper. So here  we simply present these numerical results.

\section{Conclusions}

We study the stochasticity in the galaxy-matter relation by analyzing the galaxy clustering in hyperspace of galaxy properties. Our analysis is based on the numerical simulation TNG300-1. We measure the cross power spectrum matrix $P_{ij}$ of galaxy clustering in hyperspaces of color, photometry, and color-photometry combination. We find that the principal component, namely the largest eigenmode of $P_{ij}$, is an excellent proxy for the deterministic galaxy bias. We are then able to use it as the weight to suppress stochasticity in the galaxy overdensity field and to improve the matter power spectrum reconstruction combining galaxy auto and cross clustering. For a variety of galaxy samples mimicking stage III and IV galaxy surveys, this procedure can suppress the stochasticity by a factor $2$ at $k=1h/$Mpc. We also find that the stochasticity increases monotonically with  $f_\lambda$ and $f_{\lambda^2}$. Since both can be obtained through the matrix analysis and are therefore observables, they provide useful and independent information on stochasticity. 

We find that, for the purpose of suppressing stochasticity by the described method, it is better to work in the color space or color-photometry space, instead of the photometry space. We think that this is due to the large dynamical range of $b_D$ for the color-based galaxy sub-samples. If this speculation is valid, observationally we will have extra handle on suppressing stochasticity.  We may select galaxy sub-samples with large  dynamical range of the estimated $\hat{b}_D$, and use them to suppress the stochasticity.  For the same reason, it is also useful to include more galaxy sup-samples, so the covered dynamical range can be extended.  Nevertheless, this observation is based only on a single hydro-simulation. Whether this is a universal behavior, along with the general applicability of the proposed PCA analysis in real data, are crucial for further investigation, either with more simulations or in combination with semi-analytical models of galaxy formation.

Throughout this paper we only discuss the 3D clustering. In realistic applications combining galaxy clustering and galaxy-galaxy lensing, we should work on the 2D angular clustering. Furthermore, we should include various observational issues. A significant one is the photometric redshift error and the non-zero width of redshift bins.  On one hand the projection reduces stochasticity due to suppression of small scale clustering. On the other hand, it mixes different scales and redshifts, and may induce new modes of stochasticity. To what extent the proposal described in this paper can be applied to photometric galaxy and shear catalogues, is an issue to be further investigated.

\section*{Acknowledgements}
This work is supported by the National Science Foundation of China (11621303), the National Key R\&D Program of China (2020YFC2201602) and CMS-CSST-2021-A02.
This work made use of the Gravity Supercomputer at the Department of Astronomy, Shanghai Jiao Tong University.

\section*{Data Availability}
The data underlying this article will be shared on reasonable request to the corresponding author.



\bibliographystyle{mnras}
\bibliography{main} 


\appendix

\section{ Another derivation of weighted estimator }
\label{append:another_derivation}
Instead of directly weighting the power spectra to suppress stochasticity, we may weigh  overdensity maps $\delta_{g,i}$ of each galaxy sub-samples ($i=1,\cdots N$) to form a map $\hat{\delta}_m$ of reduced stochasticity. We seek for a linear estimator of the form
\begin{equation}
    \hat{\delta}_m=\sum_i W_i \delta_{g,i} \ .
\end{equation}
The above applies in both real and Fourier space. For brevity, we restrict to the Fourier space. 
$\hat{\delta}_m$ is related to the matter overdensity $\delta_m$ by
\begin{equation}
\label{eqn:dhat}
    \hat{\delta}_m=\left(\sum_i W_i b_{D,i}\right)\delta_m+\sum_i W_i \delta_{g,i}^{S}\ .
\end{equation}
Since $b_{D,i}\propto E_i^{(1)}$ to excellent accuracy, we  then require 
\begin{equation}
\begin{split}
    \left \{
    \begin{array}{ll}
         &  \langle \hat{\delta}^2\rangle {\rm \  is\ minimalized} \\
         &  \sum_i W_i E^{(1)}_{i}=1 \\
    \end{array}
    \right .
\end{split}
\end{equation}
The solution is 
\begin{equation}
    W_i=E^{(1)}_i\ .
\end{equation}
This map then has an auto power spectrum  $P_{\hat{\delta}_m\hat{\delta}_m}=\lambda_1$ and a cross power spectrum $P_{\hat{\delta}_m \delta_m}$ with the matter map. We can then form an estimator of $P_{mm}$,
\begin{equation}
    \hat{P}_{mm}\equiv \frac{P^2_{\hat{\delta}_m \delta_m}}{P_{\hat{\delta}_m\hat{\delta}_m}}\ .
\end{equation}
This estimator is identical to Eq. \ref{eqn:Pmm_weighted}. Since $b_{D,i}\propto E_i^{(1)}$ is essentially exact, the weight $W_i$ also  suppresses the stochasticity term  (the last term in Eq. \ref{eqn:dhat}) with respect to the deterministic clustering term. But to what extend is the suppression depends on the galaxy hyperspace performing the matrix operation and weighting, as shown in the main text.


\bsp	
\label{lastpage}
\end{document}